\documentclass[12pt]{article}
\newtheorem{theorem}{Theorem}[section]
\newtheorem{remark}{Remark}

\newtheorem{lemma}{Lemma}

\usepackage{bbold}
\usepackage{caption}
\usepackage{amsmath}
\usepackage{multirow}
\usepackage{amsfonts}
\usepackage{amssymb}
\usepackage{enumitem}
\usepackage{graphicx}
\usepackage{float}
\usepackage{rotating, color}
\usepackage{undertilde}
\usepackage{theoremref}
\allowdisplaybreaks
\numberwithin{equation}{section}
\usepackage[round]{natbib}

\DeclareMathOperator*{\argmax}{argmax} 

\begin{document}

\title{ }
\author{}
\date{}
\begin{center}
{\large\bf ANALYZING RELEVANCE VECTOR MACHINES USING A SINGLE PENALTY APPROACH}\\
%MS Creative Component\\ 
Anand Dixit{\footnote[2]{Corresponding author email address: adixitstat@gmail.com}} and Vivekananda Roy\\ 
%~~Advisor : Vivekananda Roy\\ 
%POS Committee Members : Alicia Carriquiry and Mark Kaiser\\
Department of Statistics, Iowa State University
\end{center} 

\begin{abstract}
Relevance vector machine (RVM) is a popular sparse Bayesian learning model typically used for prediction. Recently it has been shown that improper priors assumed on multiple penalty parameters in RVM may lead to an improper posterior. Currently in the literature, the sufficient conditions for posterior propriety of RVM do not allow improper priors over the multiple penalty parameters. In this article, we propose a single penalty relevance vector machine (SPRVM) model in which multiple penalty parameters are replaced by a single penalty and we consider a semi Bayesian approach for fitting the SPRVM. The necessary and sufficient conditions for posterior propriety of SPRVM are more liberal than those of RVM and allow for several improper priors over the penalty parameter. Additionally, we also prove the geometric ergodicity of the Gibbs sampler used to analyze the SPRVM model and hence can estimate the asymptotic standard errors associated with the Monte Carlo estimate of the means of the posterior predictive distribution. Such a Monte Carlo standard error cannot be computed in the case of RVM, since the rate of convergence of the Gibbs sampler used to analyze RVM is not known. The predictive performance of RVM and SPRVM is compared by analyzing three real life datasets.   
\end{abstract}

 
\section{Introduction}
Let $\{(y_i, x_i):i = 1, 2, \cdot\cdot\cdot, n\}$ denote the training dataset where $y_i \in \mathcal{R}$ is the $i^{th}$ observation of the response variable and $x_i \in \mathcal{R}^p$ is the $p$ dimensional covariate vector associated with $y_i$. For such a dataset, often the objective is to come up with a function $h$, such that the response variable $y_i$ can be expressed as $y_i = h(x_i) + \epsilon_i~\forall~i = 1, 2, \cdot\cdot\cdot, n$ where $h:\mathcal{R}^p \rightarrow \mathcal{R}$ and $\{\epsilon_i\}_{i=1}^n$ are the errors. Many times, for a previously unobserved $p$ dimensional covariate vector, the function $h$ is utilized to predict its associated response variable. If $p$ is small, then the function $h$ can be estimated using the nonparametric approach of a Nadaraya-Watson type estimator. In this approach, the errors are assumed to be uncorrelated, have a zero mean and a constant variance. For higher dimensions, kernel density estimation might not work well, and hence Nadaraya Watson type estimators are not recommended when $p$ is large. Thus, in cases where $p$ is large but smaller than $n$, one can use the ordinary least squares (OLS) method to estimate the function $h$. In OLS, $h$ is estimated from a class of linear models by minimizing the quadratic loss function.  

In recent years, there is a plethora of datasets wherein $p$ is far greater than $n$. Such datasets are often referred to as high dimensional datasets. Examples of these can be found in the field of genetics, nutrition, chemical engineering etc. In such cases, the methods described before are no longer applicable. A possible solution in such cases is to use the least absolute shrinkage and selection operator (LASSO) proposed by \cite{tib::1996} that estimates the function $h$ from a class of linear models by minimizing the quadratic loss function with respect to an $L_1$ penalty. Another option is to utilize the ridge estimator proposed by \cite{hoe:ken::1970} that is similar to LASSO, but uses an $L_2$ penalty. There are other penalized regression variants of LASSO and ridge proposed in the literature (see eg. \cite{zhu:has::2005}). Bayes and empirical Bayes versions have also been developed using the connection between the penalized estimates and the posterior mode corresponding to appropriately chosen prior densities on the regression coefficients (see \cite{par:cas::2008}, \cite{kyu::2010} and \cite{roy::las}). Parameter estimation in the Bayesian models is generally carried out using Markov chain Monte Carlo (MCMC) samplers. In traditional as well as Bayesian versions, a drawback of these penalized regression methods is that the function $h$ is restricted among the class of linear models. 

If one wishes to explore a more general class of models, a common strategy is to take a reproducing kernel Hilbert space (RKHS) approach to estimate the function $h$. Such an estimate of the function $h$ was found by \cite{wah::1990} by solving the Tikhonov regularization over RKHS. This RKHS based solution allows us to reduce the complexity of the model matrix from $p$ to $n$ dimensions. This pleasing property of the RKHS based solution was utilized by \cite{tip::2001} to propose the relevance vector machine (RVM) (see also \cite{tip::2000} and \cite{bis:tip::2000}). 

RVM is a hierarchical Bayesian model in which the finite dimensional solution found by \cite{wah::1990} was utilized as the mean structure of the data model. It can be analyzed using either proper or improper priors over the hyperparameters and \cite{tip::2001} presents both cases. Assuming improper priors is fine as long as the posterior propriety has been established. Recently, \cite{dix:roy::2018} provide necessary and sufficient conditions for posterior propriety of RVM and prove that improper priors assumed by \cite{tip::2001} lead to improper posteriors. Thus, in order to conduct valid Bayesian analysis, one needs to either use proper priors or other improper priors that satisfy the sufficient conditions. For additional details about RVM and some other kernel methods see \citet{cla::book}.  

In the past, \cite{fok::2011} have attempted to implement RVM using conjugate proper priors over its hyperparameters. In that case, the full conditional distributions of the parameters involved in RVM are well known distributions which are easy to simulate from and hence can be utilized to construct an RVM Gibbs sampler. Further, for a previously unobserved $p$ dimensional covariate vector, the response variable can be predicted by utilizing the RVM Gibbs sampler iterations to produce a Monte Carlo estimate of the mean of the posterior predictive distribution. A Monte Carlo estimate should ideally be accompanied by a valid standard error estimate, so that the user is aware about the uncertainty associated with the estimate. In order to compute Monte Carlo standard errors for Markov chain samples, one needs to establish a Markov chain central limit theorem (CLT), which in turn depends on the rate of convergence of the Markov chain (see \cite{jon:hob::2001}). Currently in the literature, the rate of convergence of the Gibbs sampler implemented by \cite{fok::2011} is not known and hence the Markov chain CLT is not guaranteed. Thus, in the case of RVM, one cannot compute the standard errors associated with the Monte Carlo estimate of the mean of the posterior predictive distribution. 

\cite{mal::2005} proposed RKHS based hierarchical Bayesian classification models using both single and multiple shrinkage parameters which are also known as penalty parameters. RVM proposed by \cite{tip::2001} is a RKHS based hierarchical Bayesian regression model based on multiple penalty parameters. In this article we propose to replace these multiple penalty parameters by a single penalty parameter. We propose to name this new model as single penalty relevance vector machine (SPRVM) and analyze it using a semi Bayesian approach. In SPRVM, conjugate priors are assumed on a few parameters and since SPRVM is primarily used for prediction, other parameters are estimated using cross validation. Further, in the case of SPRVM, the posterior predictive distribution is not known in closed form, and a Gibbs sampler is implemented to produce a Monte Carlo estimate of the mean of the posterior predictive distribution. Additionally, we also prove that the Gibbs sampler implemented in the case of SPRVM converges at a geometric rate, and hence the Markov chain CLT is guaranteed. Thus, in the case of SPRVM, asymptotically valid standard error estimates can be attached to a Monte Carlo estimate of the mean of the posterior predictive distribution. This is an advantage of SPRVM over RVM. Furthermore, we show that unlike RVM, there is significant overlap in the necessary and sufficient conditions for posterior propriety of the SPRVM allowing improper priors on the penalty parameter. Finally, in the context of three real life datasets, we observe that the predictive performance of SPRVM is as good as the RVM. 

The article is structured as follows. In Section \ref{p3_s2}, we provide details about RVM and its associated Gibbs sampler. In Section \ref{p3_s3}, we introduce and provide details about SPRVM. In Section \ref{p3_s4}, we analyze some real life datasets obtained from the field of genetics, nutrition and chemical engineering to compare the predictive performance of RVM and SPRVM. In this section, we also discuss the marginal likelihood approach of estimating a few SPRVM model parameters and some concluding remarks are provided in Section \ref{p3_s5}. 
 
%Let $K$ be the $n \times (n+1)$ kernel matrix whose $i^{th}$ row is given by $K_{i}^T = \big(1, k_{\theta}(x_i, x_1), k_{\theta}(x_i, x_2), \cdot\cdot\cdot,  k_{\theta}(x_i, x_n)\big)$ where $\{k_{\theta}(x_i, x_j): i=1, 2, \cdot\cdot\cdot, n; j=1, 2, \cdot\cdot\cdot, n\}$ are the values of the reproducing kernel and $\theta$ is a kernel parameter that is typically tuned using cross validation.

\section{Relevance Vector Machine}\label{p3_s2}

Let $y = (y_1, y_2, \cdot\cdot\cdot, y_n)$ be the vector of standardized responses where $y_i \in \mathcal{R}$. Recall that $x_i \in \mathcal{R}^p$ denote the covariate vector associated with the $i^{th}$ observation. Let $K_\theta$ be the $n \times (n+1)$ kernel matrix whose $i^{th}$ row is given by $K_{\theta i}^T = \big(1, k_{\theta i1}, k_{\theta i2}, \cdot\cdot\cdot,  k_{\theta in}\big)$ where $\{k_{\theta ij}=k_{\theta}(x_i, x_j): i,j=1, 2, \cdot\cdot\cdot, n\}$ are the values of the reproducing kernel and $\theta$ is a kernel parameter that is typically tuned using cross validation. Also, let $\beta = (\beta_0, \beta_1, \cdot\cdot\cdot, \beta_n)$. Then, the RVM proposed by \cite{tip::2001} is as follows, 
\begin{subequations}
\label{eq:rvm}
\begin{align}
y|\beta, \sigma^2, \theta &\sim N_n(K_\theta \beta, \sigma^2I),\label{eq:yrvm}\\
\beta|\lambda_0, \lambda_1, \cdot\cdot\cdot, \lambda_n &\sim N_{n+1}(0, D^{-1})~~\text{with}~D = diag(\lambda_0, \lambda_1, \cdot\cdot\cdot, \lambda_n),\label{eq:brvm}\\
\pi(\lambda_i) &\propto \lambda_i^{a - 1} \exp\{{-b\lambda_i}\} ~~\forall i = 0, 1, 2, \cdot\cdot\cdot, n, \label{eq:lrvm}\\
\pi\bigg(\frac{1}{\sigma^2}\bigg) &\propto \bigg(\frac{1}{\sigma^2}\bigg)^{c - 1} \exp\bigg\{-\frac{d}{\sigma^2}\bigg\}, \label{eq:srvm}
\end{align}
\end{subequations}

\noindent where $(a, b, c, d)$ are hyperparameters that are specified by the user. In the above model, \cite{tip::2001} assumed that $1/\sigma^2$ and $\{\lambda_i\}_{i=0}^n$ are apriori independent. Further, $\beta$ and $1/\sigma^2$ are also assumed to be apriori independent. The posterior density of the parameters in RVM, indexed by $\theta$, is as follows, 
\begin{equation}\label{eq:post}
\pi(\beta, 1/\sigma^2, \lambda_0, \lambda_1, \cdot\cdot\cdot, \lambda_n|y, \theta) = \frac{f(y|\beta, \sigma^2, \theta) \pi(\beta, 1/\sigma^2, \lambda_0, \lambda_1, \cdot\cdot\cdot, \lambda_n)}{m_\theta(y)},
\end{equation}  
\noindent where $f(y|\beta, \sigma^2, \theta)$ is the data model given in \eqref{eq:yrvm}, $\pi(\beta, 1/\sigma^2, \lambda_0, \lambda_1, \cdot\cdot\cdot, \lambda_n)$ is the joint prior density obtained from \eqref{eq:brvm} - \eqref{eq:srvm} and $m_\theta(y)$ is the marginal likelihood which is also known as the normalizing constant. When the posterior density given in \eqref{eq:post} is integrated over the entire parametric space, the integral is equal to 1, provided the normalizing constant exists. Therefore, the posterior distribution is proper if and only if $m_\theta(y) < \infty$.

In Bayesian analysis, prior information available with the researchers is generally incorporated by choosing the user defined hyperparameters accordingly. In the case of RVM, the prior assumed can be either proper or improper depending upon the choice of hyperparameters and \cite{tip::2001} explored both cases. The improper prior assumed by \cite{tip::2001} can be obtained by choosing $(a, b, c, d)$ to be $(0, 0, 0, 0)$. \cite{dix:roy::2018} proved that for this choice of the user specified hyperparameters, $m_\theta(y)$ is infinity, and hence the RVM implemented by \cite{tip::2001} using improper priors is based on an improper posterior distribution. Given the posterior impropriety of RVM for the hyperparameters used by \cite{tip::2001}, we choose to implement RVM using priors that satisfy the sufficient conditions for posterior propriety derived by \cite{dix:roy::2018}. RVM is typically used for predicting the response variable say $y_{new}$ for a previously unobserved $p$ dimensional covariate vector say $x_{new}$. Such a prediction is often based on the posterior predictive distribution of the model, which is given by, 
$$f(y_{new}|y, \theta) = \int_{\mathcal{R}^{n+1}\times \mathcal{R}_{+}^{n+2}} f(y_{new}|\beta, \sigma^2, \theta)~\pi(\beta, 1/\sigma^2, \lambda_0, \lambda_1, \cdot\cdot\cdot, \lambda_n|y, \theta) d\beta~d\frac{1}{\sigma^2}~d\lambda_0~d\lambda_1~\cdot\cdot\cdot~\lambda_n.$$
\noindent Further, the mean of the above posterior predictive density can be reported as the predicted value associated with $x_{new}$ and is given by, 
\begin{equation}\label{eq:mean_predpost}
E(y_{new}|y, \theta) = K_{\theta, new}^T~\bar{\beta}_{R}
\end{equation}
where $K_{\theta, new}^T = \big(1, k_\theta(x_{new}, x_1), k_\theta(x_{new}, x_2), \cdot\cdot\cdot, k_\theta(x_{new}, x_n)\big)$ and $\bar{\beta}_{R}$ is the posterior mean of the parameter $\beta$ in the case of RVM model. 

Although posterior propriety is guaranteed for priors that satisfy the sufficient conditions for posterior propriety derived by \cite{dix:roy::2018}, the marginal likelihood is still analytically intractable, and hence the posterior density given in \eqref{eq:post} and $\bar{\beta}_{R}$ in \eqref{eq:mean_predpost} are not available in closed form. A Monte Carlo estimate for $\bar{\beta}_{R}$ can be obtained by implementing a Gibbs sampler with invariant density \eqref{eq:post}. The full conditional distributions of $(\beta, 1/\sigma^2, \lambda_0, \lambda_1, \cdot\cdot\cdot, \lambda_n)$, required to implement the Gibbs sampler are as follows:
 \begin{align}
\beta|\cdot &\sim N_{n+1}\bigg(\big(K_\theta^{T}K_\theta + D\sigma^2\big)^{-1}K_\theta^{T}y, \big(K_\theta^{T}K_\theta\frac{1}{\sigma^2} + D\big)^{-1}\bigg)\nonumber\\
\frac{1}{\sigma^2}|\cdot &\sim Gamma\bigg(\frac{n}{2} + c,~\frac{1}{2}||y - K_\theta\beta||^2 + d\bigg)\nonumber\\
\lambda_i|\cdot &\sim Gamma\bigg(a + \frac{1}{2},~\frac{\beta_i^2}{2} + b\bigg)~~\forall~i = 0, 1, 2, \cdot\cdot\cdot, n. \nonumber
\end{align}

Thus, RVM Gibbs sampler given above is a two component fixed scan sampler in which, for every iteration, $(1/\sigma^2, \{\lambda_i\}_{i=0}^n)$ is drawn given $\beta$ and then $\beta$ is drawn given the other variables. Thus, an estimate of the mean of the posterior predictive distribution is given by
\begin{equation}\label{eq:est_mean_predpost}
\hat{E}(y_{new}|y, \tilde{\theta}) = K_{\tilde{\theta}, new}^T~\hat{\bar{\beta}}_{R,M}, 
\end{equation}
where $\tilde{\theta}$ is the estimate of $\theta$ found using cross validation, $K_{\tilde{\theta}, new}^T$ is as defined previously and $\hat{\bar{\beta}}_{R,M}$ is the Monte Carlo estimate of the posterior mean of $\beta$, based on $M$ iterations of the RVM Gibbs sampler. 

The choice of $M$ depends on the Monte Carlo standard error (MCSE) associated with the estimate given in \eqref{eq:est_mean_predpost}. If the MCSE associated with \eqref{eq:est_mean_predpost} is deemed large, then it can be reduced by choosing a larger $M$. On the other hand, if the error is small, computing resources can be conserved by choosing a smaller $M$. But, since the rate of convergence of the above RVM Gibbs sampler is not known, we cannot compute the Monte Carlo standard error associated with the Monte Carlo estimate given in \eqref{eq:est_mean_predpost}. Thus, in the case of RVM, there are no guidelines for choosing a suitable $M$. Additionally, if proper priors are assumed in RVM, it requires the specification of user defined hyperparameters $(a, b, c, d)$. Specifying these hyperparameters to assume a non-informative proper prior can be challenging. Therefore, \cite{fok::2011} proposed to reduce the number of hyperparameters by assuming an extended hierarchical prior structure. The sufficient conditions for posterior propriety of RVM derived by \cite{dix:roy::2018} allow for impropriety over $1/\sigma^2$ but not over $\{\lambda_i\}_{i=0}^n$. Hence in the data analysis section of this article, for $1/\sigma^2$ we assume an improper prior, $\pi(1/\sigma^2) \propto \sigma^2$ which can be obtained by choosing $c = d = 0$, and in the case of $\{\lambda_i\}_{i=0}^n$, for the sake of implementation, we choose $a = 0.001$ and $b = 0.01$ which yields a proper Gamma prior with a mean of 0.1 and a variance of 10.

\section{Single Penalty Relevance Vector Machine}\label{p3_s3}

The improper prior assumed by \cite{tip::2001} was looked upon to be non-informative and hyperparameter free. But since it leads to an improper posterior distribution, one cannot implement RVM using that improper prior. In this section we will replace multiple penalty parameters with a single penalty parameter and simplify the prior structure to propose single penalty relevance vector machine (SPRVM). 

Let $\{(y_i, x_i): i = 1, 2, \cdot\cdot\cdot, n\}$ be the training data containing standardized responses and their corresponding covariate vectors, $\beta$ be the vector of coefficient parameters and $K_\theta$ be the $n \times (n+1)$ kernel matrix, where $y_i$, $x_i$, $\beta$ and $K_\theta$ are as defined previously in Section \ref{p3_s2}. Then we propose SPRVM as follows, 
\begin{subequations}
\label{eq:sprvm}
\begin{align}
y|\beta, \xi, \theta &\sim N(K_\theta\beta, \xi^{-1}I),\label{eq:ysprvm}\\
\beta|\lambda &\sim N(0, \lambda^{-1} I),\label{eq:bsprvm}\\
\pi(\lambda) &\propto \lambda^{a - 1} \exp\{{-b\lambda}\}\label{eq:lsprvm}
\end{align}
\end{subequations}
where $(a, b)$ are user specified hyperparameters. If a Gamma prior is assumed on $\xi$, then a Gibbs sampler can be implemented. Such an MCMC sampler does not work well in practice since the traceplot for the $\xi$ parameter reveals mixing issues. Therefore, for SPRVM, we do not assume any prior over $\xi$. For SPRVM, the posterior density of parameters $(\beta, \lambda)$, indexed by $\theta$ and $\xi$, is as follows, 
\begin{equation}\label{eq:post_sprvm}
\pi(\beta, \lambda|y, \xi, \theta) = \frac{f(y|\beta, \xi, \theta) \pi(\beta, \lambda)}{m_{\theta, \xi}(y)},
\end{equation}
where $f(y|\beta, \xi, \theta)$ is the data model given in \eqref{eq:ysprvm}, $\pi(\beta, \lambda)$ is the joint prior density following from \eqref{eq:bsprvm}-\eqref{eq:lsprvm} and $m_{\theta, \xi}(y)$ is the marginal likelihood which is given by, 
\begin{equation}\label{eq:mlik_c}
m_{\theta, \xi}(y) = \int_{\mathcal{R}^{n+1}\times \mathcal{R}_+} f(y|\beta, \xi, \theta) \pi(\beta, \lambda)~ d\beta ~ d\lambda. 
\end{equation} 

As mentioned previously in Section \ref{p3_s2}, the posterior density given in \eqref{eq:post_sprvm} is proper if and only if the marginal likelihood exists, i.e., if $m_{\theta, \xi}(y) < \infty$. For SPRVM, the necessary conditions for the posterior propriety are as follows.

\begin{theorem}\label{thm_necc}
Consider the SPRVM given in \eqref{eq:sprvm}, then, for $b=0$, which leads to the prior $\pi(\lambda) \propto \lambda^{a - 1}$, a necessary condition for the propriety of the posterior density \eqref{eq:post_sprvm} is $a \in (-(n +1)/2, 0)$.
\end{theorem}
\noindent A proof of Theorem \ref{thm_necc} is given in the Appendix B.\\ 

\noindent The improper priors that do not satisfy the above necessary conditions will lead to an improper posterior. To identify improper priors that will lead to a proper posterior, we need to derive sufficient conditions. Since the full conditional distributions of $(\beta, \lambda)$ are known, we can construct a Gibbs sampler to explore the analytically intractable posterior density \eqref{eq:post_sprvm}. The conditionals required for the implementation of the SPRVM Gibbs sampler are as follows, 
\begin{subequations}
\label{eq:fc}
\begin{align}
\beta|\cdot &\sim N_{n+1}\bigg(\big(K_\theta^{T}K_\theta + \lambda\xi^{-1} I\big)^{-1}K_\theta^{T}y, \big(K_\theta^{T}K_\theta\xi + \lambda I\big)^{-1}\bigg)\label{eq:fc_b}\\
\lambda|\cdot &\sim Gamma\bigg(\frac{n + 1}{2} + a,~\frac{\beta^T \beta}{2} + b\bigg).\label{eq:fc_l}
\end{align}
\end{subequations}

Let $\{(\beta^{(j)}, \lambda^{(j)})\}_{j=0}^\infty$ be the fixed scan two component Markov chain associated with the SPRVM Gibbs sampler. Such a Gibbs sampler is geometrically ergodic if there exists a positive real valued function $G$ and a constant $\rho \in [0, 1)$ such that, 
\begin{equation}\label{eq:ge_defn}
||P^{t}\big((\beta_0, \lambda_0), \cdot\big) - \Pi(\cdot|y)||_{TV} \le G(\beta_0, \lambda_0)\rho^t~~~~\forall~~t = 1, 2, \cdot\cdot\cdot 
\end{equation} 
\noindent where $||\cdot||_{TV}$ denotes the total variation norm, $P^{t}\big((\beta_0, \lambda_0), \cdot\big)$ denotes the probability distribution of the SPRVM Markov chain started at $(\beta_0, \lambda_0)$ after $t$ steps and $\Pi(\cdot|y)$ is the probability measure corresponding to the posterior density given in \eqref{eq:post_sprvm}. If the geometric ergodicity of the SPRVM Gibbs sampler is established, then under finite moments, a CLT is guaranteed for the posterior mean estimates of $(\beta, \lambda)$ computed using the SPRVM Gibbs sampler draws (see \cite{rob:ros::1997}). The geometric ergodicity of SPRVM Gibbs sampler defined in \eqref{eq:ge_defn} is proved in the following theorem. 
\begin{theorem}\label{thm_ge}
The SPRVM Gibbs sampler $\{(\beta^{(j)}, \lambda^{(j)})\}_{j=0}^\infty$ is geometrically ergodic if conditions (i), (ii) and (iii) given below are satisfied. 
\begin{enumerate}[label=(\roman*)]
\item Either $b > 0$ or $a < b=0$. 
\item There exists $s \in (0, 1]$ such that, 
$$\frac{\Gamma\bigg(\dfrac{n+1}{2} + a - s\bigg)}{\Gamma\bigg(\dfrac{n+1}{2} + a\bigg)}~ < ~ 2^s.$$
\item The kernel matrix $K_\theta$ defined earlier in Section \ref{p3_s2} is such that,
$$\frac{k_{\theta ij}}{k_{\theta jj}} \ne 1~\text{and}~ k_{\theta jj} \ne 0 ~ \forall i, j = 1, 2, \cdot\cdot\cdot, n~\text{and}~i \ne j.$$ 
\end{enumerate}
\end{theorem}
\noindent A proof of Theorem \ref{thm_ge} is given in the Appendix B. 
  
\begin{remark}\label{remark1}
Taking $s = 1$, condition (ii) of Theorem \ref{thm_ge} holds for $a > -(n-2)/2$. 
\end{remark}

\begin{remark}
The following are some examples of reproducing kernels typically used in sparse Bayesian learning models. 
\begin{itemize}
\item \textbf{Gaussian kernel:}
$$k_{\theta ij} = k_\theta(x_i, x_j) = \exp \Bigg\{{-\frac{||x_i - x_j||^2}{\theta^2}}\Bigg\}~\forall~i, j = 1, 2, \cdot\cdot\cdot, n,$$
\noindent where $\theta \in \mathcal{R}_+$ and $||\cdot||$ denotes the Euclidean norm.    
\item \textbf{Laplace kernel:}
$$k_{\theta ij} = k_\theta(x_i, x_j) = \exp \Bigg\{{-\frac{||x_i - x_j||}{\theta}}\Bigg\}~\forall~i, j = 1, 2, \cdot\cdot\cdot, n,$$ 
\noindent where $\theta \in \mathcal{R}_+$.
\item \textbf{Polynomial kernel:}
$$k_{\theta ij} = k_\theta(x_i, x_j) = (1 + x_i^Tx_j)^\theta~\forall~i, j = 1, 2, \cdot\cdot\cdot, n,$$ 
\noindent where $\theta \in \mathcal{N}$.
\end{itemize}
Note that for each of the above three kernels, the condition (iii) of Theorem \ref{thm_ge} will be satisfied if $x_i \neq x_j~\forall~i, j = 1, 2, \cdot\cdot\cdot, n~\text{and}~i\neq j$.
\end{remark}
Since the conditions for geometric ergodicity are sufficient for posterior propriety, a large class of improper priors guarantee posterior propriety for SPRVM. There is also a significant overlap in the necessary and sufficient conditions for posterior propriety. The necessary and sufficient conditions for posterior propriety of RVM derived by \cite{dix:roy::2018} do not have any overlap in them. In fact, the sufficient conditions in \cite{dix:roy::2018} do not allow for any prior impropriety in multiple penalty parameters of RVM. Given the sufficient conditions for posterior propriety of SPRVM, we propose to assume the following improper prior on the penalty parameter $\lambda$, 
\begin{equation}\label{eq:imp_prior}
\pi(\lambda) \propto \frac{1}{\lambda^2}.
\end{equation}     
\noindent 
From Remark \ref{remark1}, for $n \ge 5$, the above improper prior satisfies the sufficient condition for posterior propriety of SPRVM. Thus, the above improper prior allows SPRVM to have a non-informative prior structure without the difficulty of specifying any hyperparameters and also leads to a proper posterior as long as $n \ge 5$. Thus, SPRVM is able to achieve the objective of specifying a non-informative improper prior which leads to a proper posterior. 

In SPRVM, majority of parameters are estimated by the Gibbs sampler given in \eqref{eq:fc} and since SPRVM is primarily used for prediction, the remaining parameters i.e. the precision parameter, $\xi$, and the kernel parameter, $\theta$, are estimated using cross validation. We also tried the approach of estimating $\xi$ and $\theta$ by optimizing the marginal likelihood, however, the prediction performance of this approach was found to be poor. Additional details and illustrations about the marginal likelihood approach can be found in section \ref{p3_s4}.   

In the case of SPRVM, prediction for the response variable say $y_{new}$ for a previously unobserved $p$ dimensional covariate vector say $x_{new}$ is based on posterior predictive distribution, which is given by
$$f(y_{new}|y, \tilde{\xi}, \tilde{\theta}) = \int_{\mathcal{R}^{n+1}\times \mathcal{R}_+} f(y_{new}|\beta, \tilde{\xi}, \tilde{\theta})~\pi(\beta, \lambda|y, \tilde{\xi}, \tilde{\theta})~ d\beta~d\lambda,$$
\noindent where $\tilde{\xi}$ and $\tilde{\theta}$ are the estimates of $\xi$ and $\theta$ found using cross validation.  

\noindent As observed in the case of RVM, the estimate of the mean of the above posterior predictive distribution which is reported as the predicted response corresponding to $x_{new}$ is given by
\begin{equation}\label{eq:m_pp_est}
\hat{E}(y_{new}|y, \tilde{\xi}, \tilde{\theta}) = K_{\tilde{\theta}, new}^T~\hat{\bar{\beta}}_{S,M}, 
\end{equation}   
where $K_{\tilde{\theta}, new}$ is as defined previously in \eqref{eq:mean_predpost} and $\hat{\bar{\beta}}_{S,M}$ is the estimate of the posterior mean of $\beta$ found by $\sum_{j=1}^M \beta^{(j)}/M$ where $\beta^{(j)}$'s are samples from the SPRVM Gibbs sampler given in \eqref{eq:fc}.  

From Theorem \ref{thm_ge}, we know that SPRVM Gibbs sampler converges at a geometric rate. Therefore, using Theorem \ref{thm_ge} and assuming $E[\beta^T \beta | y] < \infty$, the following central limit theorem holds, 
$$\sqrt{M}\bigg(\hat{\bar{\beta}}_{S,M} - \bar{\beta}_{S}\bigg) \rightarrow N(0, \Sigma)~~~~\text{as}~~~~ M \rightarrow \infty,$$  
where $\bar{\beta}_{S}$ is the posterior mean of $\beta$ in the case of SPRVM model and $\Sigma$ is the asymptotic covariance matrix. If the posterior mean estimate i.e. $\hat{\bar{\beta}}_{S,M}$ could be based on $M$ iid observations, then $\Sigma$ can be easily estimated using sample covariance matrix. But since $\hat{\bar{\beta}}_{S,M}$ is based on $M$ draws from the SPRVM Gibbs sampler, the draws are correlated and hence estimating $\Sigma$ is challenging. In the case of geometrically ergodic Markov chains, consistent batch means and spectral variance estimators for $\Sigma$ can be derived (see eg. \cite{vat:bio}) and these estimators are available in the mcmcse R package contributed by \cite{vat::package}. In the case of SPRVM, the estimate of the standard error associated with the Monte Carlo estimate in \eqref{eq:m_pp_est} is given by, 
\begin{equation}\label{eq:se}
\widehat{SE}(K_{new}^T~\hat{\bar{\beta}}_{S,M}) = \sqrt{K_{new}^T (\hat{\Sigma}/M) K_{new}},
\end{equation}
where $\hat{\Sigma}$ is a consistent estimator of $\Sigma$. Thus, in SPRVM, we can provide a Monte Carlo estimate of the mean of the posterior predictive distribution along with a valid estimate of its standard error. 

\section{Data Analysis}\label{p3_s4}

In order to compare the predictive performance of RVM and SPRVM, we implement these two methods on high dimensional datasets in the field of genetics, nutrition and chemical engineering. For each dataset, we split the dataset into training and testing sets. The model is fitted on the training set, and the testing set is utilized to compute the root mean squared prediction error. For both the methods we use the Gaussian kernel. For RVM, the kernel parameter $\theta$ and for SPRVM, the precision parameter, $\xi$, and the kernel parameter, $\theta$, are tuned by conducting a 10 fold cross validation. The average root mean squared prediction error (RMSPE) is computed based on 20 random splitting of the datasets into training and testing sets of size $n_\star$ and $(n - n_\star) = 10$, respectively. The details of the three datasets are as follows:

\noindent \textbf{Gene dataset:} In order to study the genetics of mice population, an experiment was conducted by \cite{lan::2006}. For the experiment, a total of $n = 60$ mice were available. Among those 60 mice, 31 were females and 29 were males. From each mouse, genetic information corresponding to 22575 genes was collected. Several physiological phenotypes were also collected. We will attempt to predict the physiological phenotype named stearoyl-CoA desaturase (SCD1) using the genetic and gender $(p = 22576)$ information available. This dataset was analyzed in the past by \cite{zha::2009} and \cite{bon:rei::2012}. It can be accessed at http://www.ncbi.nlm.nih.gov/geo; accession number GSE3330.         

\noindent \textbf{Gas dataset:} In recent years, chemical engineers have attempted to obtain the octane number of gasoline samples using near infrared (NIR) spectrum measurements. We will work with the gasoline dataset available in the pls R package and will attempt to predict the octane number of the gasoline sample using NIR spectrum measurements. The data was collected by \cite{kal::1997}, and the pls R package was contributed by \cite{mev::2016}. The dataset consists of 60 gasoline samples. For each sample, octane number and NIR spectra measurements from 900 nm to 1700 nm in 2nm intervals are provided in the dataset. Thus, the dataset consists of $n = 60$ observations and $p = 401$ variables. 

\noindent \textbf{Cookie dataset:} In the field of nutrition, researchers are often interested in finding out the fat content of food items. The ppls R package provides a cookie dataset which consists of data on 72 cookie dough samples. For each sample, fat content and NIR spectra measurements from 1100 nm to 2498 nm at 2 nm intervals are provided in the dataset. In this exercise, our objective will be to predict the fat content using NIR spectrum measurements. The R package ppls was provided by \cite{ppls}, and the dataset was collected by \cite{ppls_data}. This dataset was analyzed in the past by \cite{bro::2001} among others. Among the 72 observations, 2 are outliers which are often excluded from analysis. Thus, the dataset consists of $n = 70$ observations and $p = 700$ variables.    
 
For RVM and SPRVM Gibbs samplers, we run four independent chains using over dispersed starting values for 5000 iterations and assess convergence using potential scale reduction factor (PSRF) proposed by \cite{gel:rub:1992}. The PSRF values for all the variables in RVM and SPRVM were close to 1. We also investigated the corresponding traceplots and observed that the MCMC sampler was fairly stable and there were no signs of non convergence. Thus, for both RVM and SPRVM, in order to draw observations from the posterior predictive distribution, the corresponding Gibbs samplers were run for 10000 iterations out of which first 5000 were treated as burn-in. 

\begin{table}[h]
\caption{Comparing the predictive performance of RVM and SPRVM using RMSPE}
%\captionsetup{font=scriptsize}
\centering
\begin{tabular}{|c|c|c|c|}
  \hline
\textbf{Method} & \textbf{Cookie dataset} & \textbf{Gas dataset} & \textbf{Gene dataset} \\ 
  \hline

\textbf{RVM} & 0.2445 & 0.1816 & 0.6446 \\ 
\textbf{SPRVM} & 0.2379 & 0.1725 & 0.5852 \\ 
\textbf{SPRVM-ML} & 0.3675 & 0.1668 & 0.6137 \\ 

   \hline
\end{tabular}
\label{tab:c1_tab}
\end{table}

In Table \ref{tab:c1_tab} we observe that the predictive performance of SPRVM is either similar or slightly better than that of RVM. The advantage of SPRVM over RVM is that, we can provide an asymptotically valid standard error estimate along with the Monte Carlo estimate of the mean of the posterior predictive distribution. To provide an illustration, for the gas dataset, consider an out of sample observation in which $y_{new} = 1.0237$. The Monte Carlo estimate of mean of the posterior predictive distribution for that observation was found to be $0.9589$ in the case of RVM and $0.9468$ in the case of SPRVM. Further, in the case of SPRVM, using \eqref{eq:se}, the associated Monte Carlo standard error was found to be $0.0022$. Thus, in the case of SPRVM, we are able to quantify the uncertainty associated with our Monte Carlo estimate.\\ 

\noindent {\bf Estimating ($\theta$, $\xi$):} For SPRVM, so far we have discussed estimating the precision parameter, $\xi$, and the kernel parameter, $\theta$, using cross validation. Another approach to estimating these parameters is by optimizing the marginal likelihood. The marginal likelihood for SPRVM is given in \eqref{eq:mlik_c}. Since the data model and the prior on $\beta$ are both normal, we can integrate it out and a simplified version of marginal likelihood is then given by, 

\begin{equation}\label{eq:mlik_sim}
m_{\theta, \xi}(y) = \int_{\mathcal{R}_+} f(y|\lambda, \xi, \theta) \pi(\lambda)~ d\lambda,
\end{equation}

\noindent where,
\begin{equation}\label{eq:simp_mlik} 
f(y|\lambda, \xi, \theta) = \dfrac{\xi^{-1/2}}{(2\pi)^{n/2}}~\lambda^{(n+1)/2}~|K^TK + \lambda\xi^{-1} I|^{-1/2}~\exp\bigg\{-\frac{1}{2}y^T\big(\xi^{-1} I + \lambda^{-1}KK^T\big)^{-1}y\bigg\}
\end{equation}
\noindent and $\pi(\lambda)$ is as given in \eqref{eq:imp_prior}. The estimate of $\xi$ and $\theta$ found by optimizing \eqref{eq:mlik_sim} is then given by, ($\hat{\xi}$, $\hat{\theta}$) = $\argmax m_{\theta, \xi}(y)$. 

\noindent To assess the predictive performance of the above approach, we implement it on the datasets mentioned earlier. In Table \ref{tab:c1_tab}, for the Cookie dataset, the predictive performance of the SPRVM marginal likelihood (SPRVM-ML) approach is significantly worse than that of the SPRVM cross validation approach (SPRVM). This indicates that optimizing the marginal likelihood need not be optimal from a prediction standpoint. Hence, for SPRVM, we recommend taking a cross validation approach to estimating the precision parameter, $\xi$, and the kernel parameter, $\theta$. 

\section{Conclusion}\label{p3_s5}
In this article we have proposed to analyze RVM using a single penalty parameter instead of multiple penalty parameters. The single penalty relevance vector machine (SPRVM) model was analyzed using a semi Bayesian approach. In the case of SPRVM, the sufficient conditions for posterior propriety allow for several improper priors over the penalty parameter. Currently in the literature, improper prior is not allowed over any of the penalty parameters in RVM. Additionally, we also prove the geometric ergodicity of the Gibbs sampler used to analyze the SPRVM model, and hence using the Markov chain CLT, we can calculate standard errors associated with the Monte Carlo estimate of the mean of the posterior predictive distribution. Such a measure of uncertainty cannot be computed in the case of RVM since the rate of convergence of the RVM Gibbs sampler is currently not known in the literature. Thus, the SPRVM model proposed in this article has advantages over the RVM.

\section{Appendix A: Some Useful Lemmas}

\noindent\textbf{Notation:} From here on, to simplify notations we will drop the subscript $\theta$ and write $K_\theta$ as $K$ and $k_{\theta ij}$ as $k_{ij}~\forall~i, j = 1, 2, \cdot\cdot\cdot, n$.
   
\begin{lemma}\label{l1}
Let $y$ be an $n$ dimensional vector, $K$ be an $n \times (n+1)$ matrix and $s > 0$. There exists a finite constant $Q$ depending on $y$ and $K$ such that
$$\bigg(y^TK(K^TK + \lambda \xi^{-1} I)^{-2}K^T y\bigg)^s \le Q.$$
\end{lemma} 
\noindent\textbf{Proof:} By definition, $ K^T K = \sum_{i=1}^n t_i t_i^T$ where $t_i^T$ is the $i^{th}$ row of the matrix $K$ for all $i = 1, 2, \cdot\cdot\cdot, n$. The vector $y$ can be expressed as, $y = \sum_{j=1}^n b_j e_j$ where for each $j$, $b_j \in \mathcal{R}$ and $e_j$ is the $j^{th}$ unit vector with $1$ in the $j^{th}$ place and 0 everywhere else, $j = 1, 2, \cdot\cdot\cdot, n$. Therefore,   
\begin{align}
y^TK(K^TK + \lambda \xi^{-1} I)^{-2}K^T y &= \bigg(\sum_{i=1}^n b_i e_i^T K\bigg) (K^TK + \lambda \xi^{-1} I)^{-2} \bigg(\sum_{j=1}^n b_j K^T e_j\bigg)\nonumber\\
&=\sum_{i=1}^n\sum_{j=1}^n b_i b_j t_i^T \bigg(\sum_{k=1}^n t_k t_k^T + \lambda \xi^{-1} I\bigg)^{-2} t_j.\label{eq:l1_eq0}
\end{align}
 \noindent Using Lemma 3 of \cite{kha:hob::2011}, $t_i^T \bigg(\sum_{k=1}^n t_k t_k^T + \lambda \xi^{-1} I\bigg)^{-2} t_i \le Q_i~ \forall~i = 1, 2, \cdot\cdot\cdot, n,$ where $\{Q_i: i = 1, 2, \cdot\cdot\cdot, n\}$ are constants that depends on $n, t_1, t_2, \cdot\cdot\cdot, t_n.$ 

\noindent By Cauchy-Schwartz inequality, 
\begin{align}
\bigg[t_i^T \bigg(\sum_{k=1}^n t_k t_k^T + \lambda \xi^{-1} I\bigg)^{-2} t_j\bigg]^2 &\le \bigg[t_i^T \bigg(\sum_{k=1}^n t_k t_k^T + \lambda \xi^{-1} I\bigg)^{-2} t_i\bigg]~\bigg[t_j^T \bigg(\sum_{k=1}^n t_k t_k^T + \lambda \xi^{-1} I\bigg)^{-2} t_j\bigg] \nonumber\\ 
&\le Q_i~Q_j~~~~\forall~i,j = 1, 2, \cdot\cdot\cdot, n.\label{eq:l1_eq1} 
\end{align}

\noindent The proof follows from \eqref{eq:l1_eq0} and \eqref{eq:l1_eq1} with $Q = \bigg(\sum_{i=1}^n \sum_{j=1}^n |b_i~b_j|~\sqrt{Q_iQ_j}\bigg)^s$.

\begin{lemma}\label{l2}
Suppose $K$ is a $n \times (n+1)$ kernel matrix defined previously in Section \ref{p3_s2} that satisfies condition $(iii)$ of Theorem \ref{thm_ge}. Then, $K$ is a full row rank matrix. 
\end{lemma} 
\noindent \textbf{Proof:} Let $\alpha_i \in \mathcal{R}$ for all $i = 1, 2, \cdot\cdot\cdot, n$.
\noindent We need to show that $\alpha_1=\alpha_2= \cdot\cdot\cdot = \alpha_n = 0$ is the only solution that satisfies the following equations, 
\begin{align}
\sum_{i=1}^n \alpha_i &= 0\label{eq:l2_eq1}\\
\sum_{i=1}^n \alpha_i k_{ij} &= 0~~\forall~j =  1, 2, \cdot\cdot\cdot, n.\label{eq:l2_eq2} 
\end{align}

\noindent Using \eqref{eq:l2_eq1} we get, 
\begin{equation}\label{eq:l2_eq3}
\alpha_j = -\sum_{\substack{i=1\\i\ne j}}^n \alpha_i ~~\forall~j =  1, 2, \cdot\cdot\cdot, n.
\end{equation} 

\noindent Using \eqref{eq:l2_eq2} we get, 
\begin{equation}\label{eq:l2_eq4}
\alpha_j = \frac{-1}{k_{jj}}\sum_{\substack{i=1\\i\ne j}}^n \alpha_i k_{ij}~~\forall~j =  1, 2, \cdot\cdot\cdot, n.
\end{equation} 
\noindent Further, using \eqref{eq:l2_eq3} and \eqref{eq:l2_eq4}, we get, 
\begin{equation}\label{eq:l2_eq5}
\sum_{\substack{i=1\\i\ne j}}^n \alpha_i \bigg(1 - \frac{k_{ij}}{k_{jj}}\bigg) = 0~~\forall~j =  1, 2, \cdot\cdot\cdot, n.
\end{equation}

\noindent Using condition $(iii)$ of Theorem \ref{thm_ge}. $\exists~ \gamma_1 \in \mathcal{R} - \{0\}$ and $\gamma_2 \in \mathcal{R} - \{0\}$ such that, 
\begin{equation}\label{eq:l2_eq6}
\gamma_1\sum_{\substack{i=1\\i\ne j}}^n \alpha_i~ \le~ 
\sum_{\substack{i=1\\i\ne j}}^n \alpha_i \bigg(1 - \frac{k_{ij}}{k_{jj}}\bigg)~\le ~ 
\gamma_2\sum_{\substack{i=1\\i\ne j}}^n \alpha_i~~\forall~j =  1, 2, \cdot\cdot\cdot, n.  
\end{equation}
\begin{itemize}

\item \textbf{Case 1:} If $\gamma_1, \gamma_2 > 0$ or $\gamma_1, \gamma_2 < 0$, then from \eqref{eq:l2_eq5} and \eqref{eq:l2_eq6} we have,  
\begin{equation}\label{eq:l2_eq7}
\sum_{\substack{i=1\\i\ne j}}^n \alpha_i = 0~~\forall~j =  1, 2, \cdot\cdot\cdot, n.   
\end{equation} 
\noindent Using \eqref{eq:l2_eq1} and \eqref{eq:l2_eq7}, $\alpha_1=\alpha_2= \cdot\cdot\cdot = \alpha_n = 0$ is the only possible solution.

\item \textbf{Case 2:} If $\gamma_1 > 0~\text{and}~\gamma_2 < 0$, then from \eqref{eq:l2_eq5} and \eqref{eq:l2_eq6} we have,
\begin{equation}\label{eq:l2_eq8}
\sum_{\substack{i=1\\i\ne j}}^n \alpha_i \le 0~~\forall~j =  1, 2, \cdot\cdot\cdot, n.   
\end{equation} 
\noindent Using \eqref{eq:l2_eq1} and \eqref{eq:l2_eq8}, $\alpha_1=\alpha_2= \cdot\cdot\cdot = \alpha_n = 0$ is the only possible solution.

\item \textbf{Case 3:} If $\gamma_1 < 0~\text{and}~\gamma_2 > 0$, then from \eqref{eq:l2_eq5} and \eqref{eq:l2_eq6} we have, 
\begin{equation}\label{eq:l2_eq9}
\sum_{\substack{i=1\\i\ne j}}^n \alpha_i \ge 0~~\forall~j =  1, 2, \cdot\cdot\cdot, n.   
\end{equation} 
\noindent Using \eqref{eq:l2_eq1} and \eqref{eq:l2_eq9}, $\alpha_1=\alpha_2= \cdot\cdot\cdot = \alpha_n = 0$ is the only possible solution.
\end{itemize}
\noindent Thus, combining Case 1, Case 2 and Case 3, $K$ is a full row rank matrix, i.e. $rank(K) = n$.\\ 

\begin{lemma}\label{l3}
Suppose $K$ is a kernel matrix as defined in Section \ref{p3_s2} that satisfies condition $(iii)$ of Theorem \ref{thm_ge} and $s \in (0, 1]$ then, 
$$\bigg[tr\bigg((K^TK \xi + \lambda I)^{-1}\bigg)\bigg]^s \le \xi^{-s}\bigg[tr\bigg((K^TK)^{+}\bigg)\bigg]^s + \lambda^{-s},$$
where $(K^TK)^{+}$ denotes the Moore Penrose inverse of $K^TK$.      
%$P_{K^TK}$ denotes orthogonal projection onto column space of $K^TK$ 
\end{lemma}      
\noindent \textbf{Proof:} Let $O\Psi O^T$ be the spectral decomposition of $K^T K$ where $O$ is an orthogonal matrix such that its columns $\{o_i\}_{i=1}^{n+1}$ are eigenvectors of $K^TK$ and $\Psi = diag(\psi_1, \psi_2, \cdot\cdot\cdot, \psi_{n+1})$ is a diagonal matrix whose diagonal elements are eigenvalues of $K^TK$. Then,  
\begin{equation}\label{eq:l3_eq1}
\bigg(K^T K \xi + \lambda I\bigg)^{-1} = O \bigg(\Psi \xi + \lambda I\bigg)^{-1} O^T.
\end{equation}     
\noindent As in \cite{abr:hob::2017}, let $\Psi^{+}$ be a $(n+1)\times (n+1)$ diagonal matrix whose $i^{th}$ diagonal element is given by, 
$$\psi_i^{+} = \psi_i^{-1}(1 - I_{\{0\}}(\psi_i))~~\forall~~i = 1, 2, \cdot\cdot\cdot, n+1.$$
\noindent Further, 
\begin{equation*}
\bigg(\psi_i \xi + \lambda\bigg)^{-1} \le~ \xi^{-1}\psi_i^{+} + \lambda^{-1} I_{\{0\}}(\psi_i^{+})~~\forall~~i = 1, 2, \cdot\cdot\cdot, n+1.
\end{equation*}
So, 
\begin{equation}\label{eq:l3_eq2}
 \bigg(\Psi \xi + \lambda I\bigg)^{-1} \le~\xi^{-1} \Psi^{+} + \lambda^{-1}(I - P_{\Psi})
\end{equation}                  
where $P_{\Psi}$ is a $(n+1) \times (n+1)$ diagonal matrix whose $i^{th}$ diagonal element is $1 - I_{\{0\}}(\psi_i)$.
\noindent Using \eqref{eq:l3_eq1} and \eqref{eq:l3_eq2}, we get,
\begin{align}
\bigg(K^T K \xi + \lambda I\bigg)^{-1} &\le \xi^{-1}~O\Psi^{+}O^T + \lambda^{-1}~O(I - P_\Psi)O^T\nonumber\\
&=\xi^{-1}~(K^TK)^{+} + \lambda^{-1}~O(I - P_\Psi)O^T\label{eq:l3_eq3}. 
\end{align}

\noindent Let $\tilde{O}$ be submatrix of $O$ consisting of columns $\{o_i\}_{i \in \mathcal{A}}$ where $\mathcal{A} = \{i \in \{1, 2, \cdot\cdot\cdot, n+1\}: \psi_i > 0\}$ then,  
$$OP_\Psi O^T = \sum_{i \in \mathcal{A}} o_i o_i^T = \tilde{O}\tilde{O}^T.$$

\noindent Further, $\tilde{O}\tilde{O}^T$ is an orthogonal projection onto $K^TK$ since $\{o_i\}_{i \in \mathcal{A}}$ forms an orthogonal basis for the column space of $K^T K$. Therefore, 
\begin{equation}\label{eq:l3_eq4}
O(I - P_\Psi) O^T = I - P_{K^TK}, 
\end{equation} 
\noindent where $P_{K^TK}$ denotes orthogonal projection onto column space of $K^T K$.
    
\noindent Using \eqref{eq:l3_eq3} and \eqref{eq:l3_eq4} and since $s \in (0, 1],$
\begin{align} 
\bigg(K^T K \xi + \lambda I\bigg)^{-1} &\le
\xi^{-1}~(K^T K)^{+} + \lambda^{-1}~(I - P_{K^TK})\nonumber\\
\therefore \bigg[tr\bigg((K^TK \xi + \lambda I)^{-1}\bigg)\bigg]^s &\le 
\xi^{-s}\bigg[tr\bigg((K^TK)^{+}\bigg)\bigg]^s + \lambda^{-s}~\bigg[tr(I - P_{K^TK})\bigg]^s. \label{eq:l3_eq5}
\end{align}

\noindent Further, using Lemma 2, 
\begin{equation}\label{eq:l3_eq6} 
tr(I - P_{K^T K}) = tr(I) - tr(P_{K^T K}) = (n+1) - rank(K) = 1. 
\end{equation}
\noindent Using \eqref{eq:l3_eq5} and \eqref{eq:l3_eq6}, we get, 
$$\bigg[tr\bigg((K^TK\xi + \lambda I)^{-1}\bigg)\bigg]^s \le \xi^{-s}\bigg[tr\bigg((K^TK)^{+}\bigg)\bigg]^s + \lambda^{-s}.$$
\noindent Hence proved. 

\begin{lemma}\label{l4}
Consider the following integral, 
$$\int_{\mathcal{R}_+} \frac{t^{-(a+1)}}{(g + t)^{(n+1)/2}}dt,$$
where $g$ and $a$ are constants. The above integral is finite iff $a \in (-(n+1)/2, 0)$.  
\end{lemma}
\noindent\textbf{Proof:} Suppose $t=g\tan^2\omega$, then the above integral becomes, 
$$2g^{-(a + (n+1)/2)} \int_{0}^{\pi/2} \frac{(\tan^2 \omega)^{-(a+1)}}{(sec^2\omega)^{(n+1)/2}} \tan\omega~\sec^2\omega~ d\omega.$$
\noindent Let $z = \sec^2\omega$, then the above integral becomes, 
$$g^{-(a + (n+1)/2)} \int_{1}^{\infty} (z - 1)^{-(a+1)}~z^{-((n+1)/2)}~dz.$$ 
\noindent The above integral is finite iff $a \in (-(n+1)/2, 0)$. Hence proved.\\ 

\section{Appendix B: Proof of Theorems}

\noindent\textbf{Proof of Theorem \ref{thm_necc}}\\
\noindent From \eqref{eq:mlik_sim}, 
$$m_{\theta, \xi}(y) = \int_{\mathcal{R}_+} f(y|\lambda, \xi, \theta) \pi(\lambda)~ d\lambda.$$

\noindent Using \eqref{eq:simp_mlik},
\begin{align}\label{eq:th1_eq1}
m_{\theta, \xi}(y) &= \int_{\mathcal{R}_+} \dfrac{\xi^{-1/2}}{(2\pi)^{n/2}}~\lambda^{(n+1)/2}~|K^TK + \lambda\xi^{-1} I|^{-1/2}~\exp\bigg\{-\frac{1}{2}y^T\big(\xi^{-1} I + \lambda^{-1}KK^T\big)^{-1}y\bigg\}~\lambda^{a - 1} d\lambda \nonumber\\
&\ge \dfrac{\xi^{-1/2}}{(2\pi)^{n/2}}  \exp\bigg\{-\frac{\xi}{2}y^Ty\bigg\} \int_{\mathcal{R}_+} \frac{\lambda^{a-1}}{\bigg(\xi^{-1}+ \dfrac{\psi_{max}}{\lambda}\bigg)^{\frac{n+1}{2}}}~d\lambda, \nonumber
\end{align} 

\noindent where $\psi_{max}$ is the maximum eigenvalue of $K^T K$. Using the transformation $t = 1/\lambda$, the above integral becomes, 
$$m_{\theta, \xi}(y) \ge \dfrac{\xi^{-1/2}}{(2\pi)^{n/2}}  \exp\bigg\{-\frac{\xi}{2}y^Ty\bigg\} \bigg[\frac{1}{\psi_{max}}\bigg]^{(n+1)/2}\int_{\mathcal{R}_+}\frac{t^{-(a+1)}}{\bigg(\dfrac{\xi^{-1}}{\psi_{max}} + t\bigg)^{\frac{n+1}{2}}}dt.$$

\noindent Using Lemma \ref{l4}, the above integral is finite iff $a \in (-(n+1)/2, 0)$. Hence proved.\\  

\noindent\textbf{Proof of Theorem \ref{thm_ge}}\\
\noindent Since, SPRVM Gibbs sampler is a two block Gibbs sampler, the two sub-chains $\{\beta^{(j)}\}_{j=0}^\infty$ and $\{\lambda^{(j)}\}_{j=0}^\infty$ are themselves Markov chains. Further, the rate of convergence of the three chains $\{\beta^{(j)}, \lambda^{(j)}\}_{j=0}^\infty$, $\{\beta^{(j)}\}_{j=0}^\infty$ and $\{\lambda^{(j)}\}_{j=0}^\infty$ is the same (see \cite{rob:ros::2001}). Therefore, if we prove the geometric ergodicity of one of the chains, it holds for all the three chains. We will work with the $\{\lambda^{(j)}\}_{j=0}^\infty$ chain. The Markov transition density associated with the $\{\lambda^{(j)}\}_{j=0}^\infty$ chain is given by, 
$$p_{l}(\tilde{\lambda}|\lambda) = \int_{\mathcal{R}^{n+1}} \pi(\tilde{\lambda}|\beta, y)~\pi(\beta|\lambda, y)~d\beta,$$ 
where $\pi(\tilde{\lambda}|\beta, y)$ is the density corresponding to the full conditional distribution given in \eqref{eq:fc_l} and $\pi(\beta|\lambda, y)$ is the density of the full conditional distribution given in \eqref{eq:fc_b}.

\noindent We define the drift function as follows, 
\begin{equation}\label{eq:t_0}
v(\tilde\lambda) = \tilde{\lambda}^m + \tilde{\lambda}^{-s}, 
\end{equation}
\noindent where $m \in (0, 1)$ is a positive constant that is determined in the proof and $s\in (0, 1]$ is a constant that satisfies condition $(ii)$ in Theorem \ref{thm_ge}. \\
\noindent Since the above drift function is unbounded off compact sets and $\{\lambda^{(j)}\}_{j=0}^\infty$ is a Feller chain, geometric ergodicity of the $\{\lambda^{(j)}\}_{j=0}^\infty$ chain is established by proving the following drift condition (see \cite{mey:twe::1993}), 
$$E[v(\tilde\lambda)|\lambda] = \int_{\mathcal{R}_+} v(\tilde{\lambda})~p_{l}(\tilde{\lambda}|\lambda)~d\tilde\lambda \le L + \rho v(\lambda)$$ 
\noindent where $L > 0$ and $\rho \in (0, 1)$ are finite constants.\\
\noindent Note that, 
\begin{equation}\label{eq:d_exp}
E[v(\tilde\lambda)|\lambda] = E\big[E[v(\tilde\lambda)|\beta]|\lambda\big].
\end{equation}
\noindent We start with the inner expectation in \eqref{eq:d_exp}. Also, first consider $b > 0$, 
\begin{equation}\label{eq:n1}
E[\tilde{\lambda}^m|\beta] = \dfrac{\Gamma(a + m + \frac{n+1}{2})}{\Gamma(a + \frac{n+1}{2})} \bigg(\dfrac{\beta^T\beta}{2} + b\bigg)^{-m}\le \dfrac{\Gamma(a + m + \frac{n+1}{2})}{\Gamma(a + \frac{n+1}{2})}~ b^{-m}.
\end{equation}
\noindent Now we consider the outer expectation in \eqref{eq:d_exp}. From \eqref{eq:n1} we get,  
\begin{equation}\label{eq:t_1}
E[\tilde{\lambda}^m|\lambda] \le \dfrac{\Gamma(a + m + \frac{n+1}{2})}{\Gamma(a + \frac{n+1}{2})}~ b^{-m}.
\end{equation}
\noindent Next, 
\begin{equation}\label{eq:n2}
E[\tilde{\lambda}^{-s}|\beta] = \dfrac{\Gamma(a - s + \frac{n+1}{2})}{\Gamma(a + \frac{n+1}{2})} \bigg(\dfrac{\beta^T\beta}{2} + b\bigg)^{s} \le \dfrac{\Gamma(a - s + \frac{n+1}{2})}{\Gamma(a + \frac{n+1}{2})} \bigg(\dfrac{(\beta^T\beta)^s}{2^s} + b^s\bigg).\\
\end{equation}

\noindent From \eqref{eq:n2}, using Lemma \ref{l1} and Lemma \ref{l3} we have, 
\begin{align}
E[\tilde{\lambda}^{-s}|\lambda] &\le \dfrac{\Gamma(a - s + \frac{n+1}{2})}{\Gamma(a + \frac{n+1}{2})}~ \bigg(\dfrac{1}{2^s} E[(\beta^T\beta)^s|\lambda] + b^s\bigg)\nonumber\\ 
&\le \dfrac{\Gamma(a - s + \frac{n+1}{2})}{\Gamma(a + \frac{n+1}{2})}~ \bigg(\dfrac{1}{2^s} \{E[(\beta^T\beta)|\lambda]\}^s + b^s\bigg)\nonumber\\
&\le \dfrac{\Gamma(a - s + \frac{n+1}{2})}{\Gamma(a + \frac{n+1}{2})}\bigg\{\frac{1}{2^s}\bigg[y^TK(K^TK + \lambda \xi^{-1} I)^{-2}K^T y + tr\bigg((K^TK\xi + \lambda I)^{-1}\bigg)\bigg]^s + b^s\bigg\}\nonumber\\
&\le \dfrac{\Gamma(a - s + \frac{n+1}{2})}{\Gamma(a + \frac{n+1}{2})}\frac{1}{2^s}\bigg[Q + (2b)^s + \xi^{-s}\bigg(tr\bigg((K^TK)^+\bigg)\bigg)^s + \lambda^{-s} \bigg]\nonumber\\
&\le L_0 + \dfrac{\Gamma(a - s + \frac{n+1}{2})}{\Gamma(a + \frac{n+1}{2})}\frac{1}{2^s} \lambda^{-s} \label{eq:t_2}
\end{align}
\noindent where $L_0 = \dfrac{\Gamma(a - s + \frac{n+1}{2})}{\Gamma(a + \frac{n+1}{2})}\dfrac{1}{2^s}\bigg[Q + (2b)^s + \xi^{-s}\bigg(tr\bigg((K^TK)^+\bigg)\bigg)^s\bigg].$\\

\noindent Using \eqref{eq:t_0}, \eqref{eq:t_1} and \eqref{eq:t_2}, we get,
$$E[v(\tilde{\lambda})|\lambda] \le L_1 + \rho_0~ v(\lambda)$$
\noindent where 
\[
L_1 = L_0 + \dfrac{\Gamma(a + m + \frac{n+1}{2})}{\Gamma(a + \frac{n+1}{2})}~ b^{-m}~\text{and}~\rho_0 = \dfrac{\Gamma(a - s + \frac{n+1}{2})}{\Gamma(a + \frac{n+1}{2})}\dfrac{1}{2^s}
\]
are finite constants. Further, using condition $(ii)$ of Theorem \ref{thm_ge}, $\rho_0 \in (0, 1)$. Thus proving geometric ergodicity of the SPRVM Gibbs sampler for $b > 0$. 

\noindent Now consider $b = 0$ and $a < 0$. Let $\Sigma_\beta$ denote the covariance matrix of $\beta|\lambda, y$ i.e.
$$\Sigma_\beta = (K^T K \xi + \lambda I )^{-1}~~\implies~~\Sigma_\beta^{-1} = K^T K \xi + \lambda I.$$
\noindent As defined in Lemma \ref{l3}, let $O\Psi O^T$ be spectral decomposition of $K^T K$ where $\Psi = diag(\psi_1, \psi_2, \cdot\cdot\cdot, \psi_{n+1})$. Also, let $\psi_{max} = max\{\psi_1, \psi_2, \cdot\cdot\cdot, \psi_{n+1}\}$. Therefore,  
\begin{align}
\Sigma_\beta^{-1} &\le (\psi_{max}~ \xi + \lambda)I\nonumber\\
\implies \beta^T \Sigma_\beta^{-1} \beta &\le \beta^T (\psi_{max}~ \xi + \lambda)I \beta\nonumber\\ 
\implies (\beta^T \Sigma_\beta^{-1} \beta)^{-m} &\ge \bigg[\beta^T (\psi_{max}~ \xi + \lambda)I \beta\bigg]^{-m}.\label{eq:t_3}
\end{align} 

\noindent Now, $\beta^T \Sigma_\beta^{-1} \beta|\lambda, y$ has a non central $\chi^2$ distribution with $n+1$ degrees of freedom. Using Lemma 4 of \cite{rom:hob::2012}, for $m \in(0, 1)$, we get,
\begin{equation}\label{eq:t_4}
E[(\beta^T \Sigma_\beta^{-1} \beta)^{-m}|\lambda] \le 2^{-m}~\dfrac{\Gamma(\frac{n+1}{2} - m)}{\Gamma(\frac{n+1}{2})}.
\end{equation}
\noindent Now, using \eqref{eq:t_3} and \eqref{eq:t_4},  
\begin{align}
E[(\beta^T \beta)^{-m}|\lambda] &= (\psi_{max}~\xi + \lambda)^m E\bigg[\bigg(\beta^T (\psi_{max}~ \xi + \lambda)I \beta\bigg)^{-m}\bigg|\lambda\bigg]\nonumber\\
&\le (\psi_{max}~\xi + \lambda)^m E[(\beta^T \Sigma_\beta^{-1} \beta)^{-m}|\lambda]\nonumber\\
&\le ((\psi_{max}~\xi)^m + \lambda^m)~2^{-m}~\dfrac{\Gamma(\frac{n+1}{2} - m)}{\Gamma(\frac{n+1}{2})}.\label{eq:t_5}
\end{align}
\noindent Since, 
$$E[\tilde{\lambda}^m|\beta] = \dfrac{\Gamma(a + m + \frac{n+1}{2})}{\Gamma(a + \frac{n+1}{2})} \frac{1}{2^{-m}} E[(\beta^T\beta)^{-m}|\lambda],$$
\noindent using \eqref{eq:t_5} we have, 
\begin{equation}\label{eq:t_6}
E[\tilde{\lambda}^m|\lambda] \le \dfrac{\Gamma(a + m + \frac{n+1}{2})}{\Gamma(a + \frac{n+1}{2})} \dfrac{\Gamma(\frac{n+1}{2} - m)}{\Gamma(\frac{n+1}{2})} ((\psi_{max}~\xi)^m + \lambda^m).
\end{equation}
\noindent Using \eqref{eq:t_0}, \eqref{eq:t_2} and \eqref{eq:t_6},
$$E[v(\tilde{\lambda})|\lambda] \le \tilde{L_0} + L_1 + \rho_0 \lambda^{-s} + \rho_1 \lambda^m$$
\noindent where $\rho_0$ is as defined before and
$$\tilde{L_0} = \dfrac{\Gamma(a - s + \frac{n+1}{2})}{\Gamma(a + \frac{n+1}{2})}\dfrac{1}{2^s}\bigg[Q + \xi^{-s}\bigg(tr\bigg((K^TK)^+\bigg)\bigg)^s\bigg],$$  
$$L_1 =  \dfrac{\Gamma(a + m + \frac{n+1}{2})}{\Gamma(a + \frac{n+1}{2})}~ \dfrac{\Gamma(\frac{n+1}{2} - m)}{\Gamma(\frac{n+1}{2})} (\psi_{max}~\xi)^m,$$
$$\rho_1 =  \dfrac{\Gamma(a + m + \frac{n+1}{2})}{\Gamma(a + \frac{n+1}{2})} \dfrac{\Gamma(\frac{n+1}{2} - m)}{\Gamma(\frac{n+1}{2})}.$$ 

\noindent For $m \in (0, 1) \cap (0, -a)$, \cite{rom:hob::2012} have shown that $\rho_1 < 1$. Let $L^{\star} = \tilde{L_0} + L_1$ and $\rho\star = \max\{\rho_0, \rho_1\}$. Then for $b=0$ and $a < 0$, 
$$E[v(\tilde{\lambda})|\lambda] \le L^\star + \rho^\star v(\lambda)$$
where $L^\star$ and $\rho^\star$ are finite constants. Further, $\rho^\star \in (0, 1)$ since $\rho_0 \in (0, 1)$ and $\rho_1 \in (0, 1)$. Thus, we have proved geometric ergodicity of $\{\lambda^{(j)}\}_{j=0}^\infty$ for $b=0$ and $a < 0$. Hence proved.

\bibliographystyle{kbib}
\bibliography{Dixit_Roy_SPRVM}

\end{document}